\documentstyle[12pt]{article}
\def\mathbf{\vec}

\def\ca{\c{c}\~{a}}

\begin{document}

\centerline {\LARGE Generalized Proper-Time Approach}
\vspace{.3cm}
\centerline {\LARGE For The Case Of Broken Isospin Symmetry}
\vspace{1cm}
\centerline {\large Alexander A. Osipov\footnote{On leave from the 
            Laboratory of Nuclear Problems, JINR, Dubna, Russia}, 
            Brigitte Hiller}
\vspace{.5cm}
\centerline {\it Centro de F\'{\i}sica Te\'{o}rica, Departamento de
             F\'{\i}sica}
\centerline {\it da Universidade de Coimbra, 3004-516 Coimbra, Portugal}
\vspace{1cm}

\begin{abstract}
We present a derivation of the low-energy effective meson Lagrangian of the
Nambu -- Jona-Lasinio (NJL) model on the basis of Schwinger's proper-time 
regularization of the one-loop fermion determinant. We consider the case in
which the $SU(2)\times SU(2)$ chiral symmetry of the NJL Lagrangian is broken 
by the current quark mass matrix with $\hat{m}_u\ne\hat{m}_d$. The 
non-degeneracy of $d$ and $u$ masses destroys one of the most crucial features 
of the proper-time expansion -- the chiral-invariant structure of Seeley -- 
DeWitt coefficients. We show however that systematic resummations inside the 
proper-time expansion are still possible and derive a result which is in full 
agreement with the chiral Ward -- Takahashi identities.  
\end{abstract}

\vspace{4.0mm}
\noindent
PACS number(s): 12.39.Fe, 11.30.Rd.
\vspace{4.0mm}

\newpage
\section{Introduction}

In recent papers \cite{Osipov:2000a,Osipov:2000b} we have shown how to 
determine systematically the low-energy structure of the Nambu -- Jona-Lasinio 
(NJL) model \cite{Nambu:1961} on the basis of Schwinger's proper time   
regularization of the one-loop fermion determinant 
\cite{Schwinger:1951,DeWitt:1965}. We have considered the cases with linear and
non-linear realizations of explicitly broken $SU(2)\times SU(2)$ chiral 
symmetry. We have shown that in the presence of the explicit chiral symmetry 
breaking term in the Lagrangian, the standard definition of $\ln |\det D|$ in 
terms of a proper-time integral 
\begin{equation}
\label{logdet}
   \ln |\det D|=-\frac{1}{2}\int^\infty_0\frac{dT}{T}\rho (T,\Lambda^2)
                 \mbox{Tr}\left(e^{-TD^\dagger D}\right)
\end{equation}
modifies the explicit chiral symmetry breaking pattern of the original quark 
Lagrangian and needs to be corrected in order to lead to the fermion 
determinant whose transformation properties exactly comply with the symmetry 
content of the basic Lagrangian. The reason for the necessary modification of
the result obtained on the basis of formula (\ref{logdet}) is directly 
connected with the restrictions imposed by the chiral Ward -- Takahashi (WT)
identities.                                                              

In the present paper we extend this framework to the case in which the 
$SU(2)\times SU(2)$ chiral symmetry of the NJL Lagrangian is broken by the 
current quark mass matrix with $\hat{m}_u\ne\hat{m}_d$. The non-degeneracy of 
$d$ and $u$ masses introduces a rather large violation of isotopic spin 
conservation and destroys one of the most crucial features of the proper time 
expansion -- the chiral-invariant structure of Seeley -- DeWitt coefficients.  
It means that additionally to the already known problem of correcting the 
standard definition (\ref{logdet}) by the functional $P$ being proportional to
the current quark masses \cite{Osipov:2000a} one has to find a method which 
replaces the standard asymptotic expansion of the heat kernel in terms of 
Seeley -- DeWitt coefficients by a new expansion in terms of chiral-invariant 
structures. As we shall show, systematic resummations inside the proper-time 
expansion are possible. We describe here how calculations can be organized in 
this, at the first sight quite fuzzy, situation. The resulting series for the 
heat kernel is not anymore a proper-time expansion. We calculate its first 
terms and present the method to determine higher orders.

The spontaneous breakdown of the global $SU(2)\times SU(2)$ chiral symmetry in 
the NJL model is a consequence of the fact that the Schwinger -- Dyson
equations for the fermion propagators (``gap"-equations) have nontrivial 
solutions with $m_u\ne m_d\ne 0$. After bosonization the same equations occur 
as conditions for the minimum of the effective potential. This generates a 
corresponding redefinition of the scalar fields, from which it follows that the
part of the effective Lagrangian which is linear in the scalar fields vanishes 
(due to the gap-equations), for excitations about the real vacuum state. 
As we shall show, this immediately affects the isospin symmetry breaking part 
of the Lagrangian and breaks the chiral WT identities at this level. As a 
consequence we need to invoke new counterterms which would have not been 
required if there had not been spontaneous symmetry breakdown. These 
counterterms have a simple structure and are completely fixed by chiral 
symmetry. 
                                                             
This article presents a new systematic expansion procedure for the heat 
kernel of the one loop fermion determinant for the case $\hat{m}_u\ne
\hat{m}_d$. It differs from the methods which have already been used in the
literature in the same context, for example in papers 
\cite{Volkov:1984,Ebert:1986}. The result is also different. In Sec. II we 
discuss the Lagrangian of the NJL model and show that chiral $SU(2)\times SU(2)
$ transformations of quark fields dictate the transformation laws of the 
auxiliary bosonic fields. These collective variables are necessary to rearrange
the four-quark Lagrangian of the NJL model in an equivalent Lagrangian which is
only quadratic in the quark fields. The chiral symmetry is broken explicitly 
by the current quark mass matrix with $\hat{m}_u\ne\hat{m}_d$. We use the 
classical equations of motion for the pseudoscalar and scalar fields to rewrite
the variation of the NJL Lagrangian with respect to the action of the chiral 
group in terms of meson fields. In Sec. III we discuss in detail the chiral WT 
identities for the case under consideration. We show how one can integrate 
these identities to get the symmetry breaking part of the effective Lagrangian 
without explicit calculations of the quark determinant. In Sec. IV we calculate
the fermion determinant. We show how to define it for the case in which 
explicit symmetry breaking takes place and isospin symmetry is broken. We 
calculate the first terms of the new expansion of the heat kernel in full 
detail. We derive the corresponding correcting polynomial from the functional 
$P$ and show that it is completely fixed by the symmetry requirements. The 
effective meson Lagrangian is obtained at the end of this section. 
In Sec. V we introduce new variables for pseudoscalar and
scalar fields in order to describe the physical meson states and to get the 
meson mass spectrum. The concluding remarks are given in Sec. VI. Finally we 
show in the Appendix some technical details in our treatment of the heat kernel
exponent with different masses for $u$ and $d$ quarks.

\section{From quarks to mesons: the infinitesimal chiral transformations}

Consider the effective quark Lagrangian of strong interactions with only light
$u$ and $d$ quark fields which is invariant under a global colour $SU(N_c)$ 
symmetry
\begin{equation}
\label{enjl}
  {\cal L}=\bar{q}(i\gamma^\mu\partial_\mu -\hat{m})q
          +\frac{G}{2}[(\bar{q}\tau_a q)^2+(\bar{q}i\gamma_5\tau_a q)^2].
\end{equation}
Here $q$ is a flavor doublet of Dirac spinors for quark fields $\bar q=(\bar 
u, \bar d)$. The fermion field carries both flavor and color indices. Summation
over the color indices is implicit. For the tau $2\times 2$ matrices, $\tau_a$,
where $a=0,1,2,3$, and $\mbox{tr}(\tau_a\tau_b)=2\delta_{ab}$, we shall use the
following notation: $\tau_0=1$ and $\tau_i, (i=1,2,3)$ are the standard Pauli 
matrices. The manifest chiral symmetry breaking occurs via the current quark 
mass matrix: $\hat{m}=\mbox{diag}(\hat{m}_u,\hat{m}_d)$ where $\hat{m}_u\ne
\hat{m}_d$. Without this term the Lagrangian (\ref{enjl}) would be invariant 
under global chiral $SU(2)\times SU(2)$ symmetry. We restrict our consideration
to the case of only scalar and pseudoscalar four-quark interactions with the 
coupling constant $G$ being strong enough to create a stable vacuum state with 
a nontrivial solution of the gap-equations corresponding to spontaneous 
breakdown of the chiral $SU(2)\times SU(2)$ symmetry.

The transformation law for the quark fields is the following
\begin{equation}
\label{quark}
   \delta q=i(\alpha +\gamma_5\beta )q, \quad
   \delta\bar{q}=-i\bar{q}(\alpha -\gamma_5\beta ),
\end{equation}
where parameters of global infinitesimal $SU(2)\times SU(2)$ chiral 
transformations are chosen as $\alpha =\alpha_i\tau_i, \ \beta =\beta_i\tau_i$.
Therefore the Lagrangian ${\cal L}$ transforms according to the law
\begin{equation}
\label{sb}
   \delta {\cal L}=i\bar{q}\left([\alpha ,\hat{m}]
                  -\gamma_5\{\beta ,\hat{m}\}\right)q.
\end{equation}
It is clear that nothing must destroy this property of the model (we are not 
considering anomalies here), otherwise the content of the theory would be 
changed. In particular it means that the symmetry breaking pattern of the 
bosonized NJL Lagrangian should be the very same, being written only in terms 
of collective fields.

To make the last statement more rigorous let us consider the path integral 
representation for the generating functional $Z$ in a theory with the 
Lagrangian density (\ref{enjl}). To integrate over anticommuting c-number quark
fields in $Z$ one has to introduce color singlet collective bosonic variables 
in such a way that the action becomes bilinear in the quark fields and the 
integration becomes trivial
\begin{equation}
\label{gf1}
   Z=\int {\cal D}q{\cal D}\bar{q}{\cal D}s_a{\cal D}p_a
     \mbox{exp}\left\{i\int d^4x\left[{\cal L} 
     -\frac{1}{2G}(s_a^2+p_a^2)\right]\right\}.
\end{equation}
We suppress external sources in the generating functional $Z$ and assume 
summation over repeated flavor indices. In order not to destroy the symmetry 
of the basic quark Lagrangian ${\cal L}$ one has to require from the new 
collective variables that
\begin{equation}
\label{req}
   \delta (s_a^2+p_a^2)=0
\end{equation}
Since there are no kinetic terms for $s_a$ and $p_a$, these fields are
auxiliary. Let us replace the variables in $Z$: $s_a, p_a\rightarrow \sigma_a ,
\pi_a$,
\begin{equation}
\label{s}
   s_a=\sigma_a -\hat{m}_a+G(\bar{q}\tau_aq),
\end{equation}
\begin{equation}
\label{p}
   p_a=\pi_a -G(\bar{q}i\gamma_5\tau_aq).
\end{equation}
Here we put 
\begin{equation}
   \hat{m}_0=\frac{\hat{m}_u+\hat{m}_d}{2},
   \quad  \hat{m}_i=\delta_{i3}\frac{\hat{m}_u-\hat{m}_d}{2}. 
\end{equation}
Requirement (\ref{req}) together with the chiral transformation laws for the 
quark fields (\ref{quark}) lead to the transformation laws for the new 
collective fields:
\begin{equation}
\label{st}
   \delta\sigma =i[\alpha, \sigma -\hat{m}]-\{\beta, \pi\}, \quad
   \delta\pi =i[\alpha, \pi ]+\{\beta, \sigma -\hat{m}\},
\end{equation}
where $\pi =\pi_a\tau_a$, $\sigma =\sigma_a\tau_a$, and $\hat{m}=\hat{m}_a
\tau_a$. In this way the transformation laws of the quark fields with respect 
to the action of chiral $SU(2)\times SU(2)$ group define the transformation 
laws of the collective bosonic fields. However Eq.(\ref{st}) is not yet the 
final form which we need. In the NJL model the vacuum is not invariant 
under chiral $SU(2)\times SU(2)$ transformations. To explore the properties of 
the spontaneously broken theory one has to define new scalar fields with 
vanishing vacuum expectation values, i.e., one has to rewrite the Lagrangian 
of the theory in terms of shifted fields:
\begin{equation}
   \sigma\rightarrow\sigma +m,
\end{equation}
where 
\begin{equation}
   m=m_a\tau_a, \quad m_0=\frac{m_u+m_d}{2},
   \quad m_i=\delta_{i3}\frac{m_u-m_d}{2}, 
\end{equation}
are the masses of constituent quarks. The chiral transformation properties of 
scalar and pseudoscalar fields change correspondingly to the new ones: 
\begin{equation}
\label{stbf}
   \delta\sigma =i[\alpha, \sigma +\Delta ]-\{\beta, \pi\}, \quad
   \delta\pi =i[\alpha, \pi ]+\{\beta, \sigma +\Delta \},
\end{equation}
where $\Delta =m-\hat{m}$. This is the final form of infinitesimal chiral
transformations for collective mesonic excitations around the nontrivial vacuum
state in the NJL model. 
 
In accordance with replacements (\ref{s}) and (\ref{p}) we obtain the mixed 
form for the NJL Lagrangian which includes both quarks and collective degrees
of freedom 
\begin{equation}
\label{lq}
   {\cal L}(q,\bar{q},\sigma_a,\pi_a)=\bar{q}Dq
           -\frac{1}{2G}[(\sigma_a +\Delta_a)^2+\pi_a^2],
\end{equation} 
where 
\begin{equation}
\label{D}
   D=i\gamma^\mu\partial_\mu -m-\sigma +i\gamma_5\pi . 
\end{equation}
The Euler -- Lagrange equations for mesonic fields take the form of constraints
\begin{equation}
   \pi_a =G\bar{q}i\gamma_5\tau_a q, \quad 
   \sigma_a =-\Delta_a -G\bar{q}\tau_a q.
\end{equation}
By using them one can rewrite Eq.(\ref{sb}) in terms of collective degrees of 
freedom:
\begin{equation}
\label{msb}
   \delta {\cal L}=\frac{\hat{m}_a}{G}\delta\sigma_{a}.
\end{equation} 
One concludes that the symmetry breaking pattern of the NJL model is not
affected by the occurence of spontaneous symmetry breakdown.

The derivation of the action for collective fields $\pi_a$ and $\sigma_a$ in 
the NJL model is reduced (after integrating out the quark fields) to the 
calculation of the functional determinant of the operator $D$
\begin{equation}
\label{seff}
   S_{\mbox{coll}}=\int d^4x {\cal L_{\mbox{coll}}}=
   -i\ln\det D-\frac{1}{2G}\int d^4x
   \left[(\sigma_a +\Delta_a)^2+\pi_a^2\right].
\end{equation}
This is a great advantage of the model, for there are methods to study the 
determinant of such operators \cite{Novikov:1984,Ball:1989}. On the other hand,
however, these methods can not be directly applyed in the presence of the 
manifest chiral symmetry breaking term in the NJL Lagrangian 
\cite{Osipov:2000a,Osipov:2000b}. We have an even more complicated case now, 
because of isospin breaking. We shall consider this problem in the 
following sections.

\section{The chiral Ward -- Takahashi identities}

One can ask if it is possible to get the symmetry breaking part of the 
bosonized NJL Lagrangian without a direct evaluation of the fermion determinant
in (\ref{seff}). The answer on this question is positive and hidden in 
Eq.(\ref{msb}). Indeed, all we need to do is just to solve it. Possible 
solutions may be constructed by applying a method which is closely related to 
the one already used in \cite{Wess:1971}. The lack of a direct dependence of 
chiral transformations on the space-time coordinates simplifies our task. In 
particular the chiral WT identities can be stated directly by giving the 
variation of the Lagrangian of collective meson fields, ${\cal L}_{\mbox{coll}}
$, with respect to the infinitesimal chiral $SU(2)\times SU(2)$ 
transformations, i.e.,
\begin{equation}
\label{sbcoll} 
   \delta {\cal L}_{\mbox{coll}}=\frac{\hat{m}_a}{G}\delta\sigma_{a}.
\end{equation} 
If chiral symmetry is spontaneously broken, the infinitesimal 
transformations are given by Eq.(\ref{stbf}). It implies that the variation of 
the functional of the considered collective meson fields is described by the 
infinitesimal operator
\begin{equation}
   \hat{\delta}=2(\alpha_i X_i + \beta_i Y_i ),
\end{equation}
where the generators $X_i$ and $Y_i$ are expressed in the space of meson fields
as follows
\begin{eqnarray}
\label{gen}
   X_i&=&-\epsilon_{ijk}\left[\pi_j\frac{\delta}{\delta\pi_k}
         +(\sigma_j +\Delta_j )\frac{\delta}{\delta\sigma_k}\right], 
         \nonumber \\
   Y_i&=&(\sigma_i +\Delta_i )\frac{\delta}{\delta\pi_0} 
         +(\sigma_0 +\Delta_0 )\frac{\delta}{\delta\pi_i}
         -\pi_i\frac{\delta}{\delta\sigma_0} 
         -\pi_0\frac{\delta}{\delta\sigma_i}.
\end{eqnarray}
The definition of a functional derivative depends on which parameters are 
considered as fixed in the functions, and which are considered as variable.
Our functions $\pi_a (x)$ and $\sigma_a (x)$ have fixed 
arguments with respect to global chiral transformations, i.e., the 
variational derivative does not produce a delta function 
of space-time coordinates. In this case the generators satisfy the commutation 
relations 
\begin{eqnarray}
\label{comrel}
   \left[X_i,X_j\right]&=&\epsilon_{ijk}X_k,  \nonumber \\   
   \left[X_i,Y_j\right]&=&\epsilon_{ijk}Y_k,  \nonumber \\   
   \left[Y_i,Y_j\right]&=&\epsilon_{ijk}X_k,  
\end{eqnarray}
which are the Lie algebra representation of the chiral group.
Using the definition (\ref{gen}), we can rewrite Eq.(\ref{sbcoll}) as
\begin{eqnarray}
\label{msbxy}
   X_i{\cal L}_{\mbox{coll}}
   &=&F_i=-\frac{1}{G}\epsilon_{ijk}\sigma_j \hat{m}_k, \nonumber \\
   Y_i{\cal L}_{\mbox{coll}}
   &=&G_i=-\frac{1}{G}(\hat{m}_0\pi_i +\hat{m}_i\pi_0 ).
\end{eqnarray}
The fact that the WT identities (\ref{msbxy}) express the variation of a single
function, ${\cal L}_{\mbox{coll}}$, gives immediately strong restrictions on 
the form of this function. In addition we have integrability conditions:
\begin{eqnarray}
\label{intcond}
   X_i F_j - X_j F_i &=&\epsilon_{ijk}F_k, \nonumber \\
   Y_i G_j - Y_j G_i &=&\epsilon_{ijk}F_k, \nonumber \\
   Y_i F_j - X_j G_i &=&\epsilon_{ijk}G_k,
\end{eqnarray}
which tell us if it is possible or not to integrate Eq.(\ref{msbxy}) with the
given functions $F_i$ and $G_i$. These conditions are fulfilled in our case. 
It is clear that from the WT identities one only obtains general symmetry 
restrictions on the form of the bosonized NJL Lagrangian; in particular, 
one can use them to fix the symmetry breaking part of ${\cal L}_{\mbox{coll}}$.
Indeed, the first equation in the system (\ref{msbxy}) gives
\begin{equation}
   {\cal L}_{\mbox{coll}}
   =-\frac{\hat{m}_3\vec{\sigma}^2}{2G\Delta_3}+{\cal L}_{S}+\Omega ,
\end{equation}
where ${\cal L}_{S}$ is a chiral symmetric part of the Lagrangian, i.e.,
$X_i{\cal L}_{S}=Y_i{\cal L}_{S}=0$, and for $\Omega$ we have
\begin{eqnarray}
   X_i\Omega &=&0, \nonumber \\
   Y_i\Omega &=&-\frac{1}{G}\left(\hat{m}_0\pi_i +\hat{m}_i\pi_0 
                +\frac{\hat{m}_3}{\Delta_3}\pi_0\sigma_i\right).
\end{eqnarray}
Noting that $\sigma_0$ and $\pi_0$ are cancelled by the action of the operator
$X_i$ ($X_i\pi_0=X_i\sigma_0=0$), we derive
\begin{equation}
   \Omega = \frac{\hat{m}_0\sigma_0}{G}-\frac{\hat{m}_3\pi_0^2}{2G\Delta_3}.
\end{equation}
Thus
\begin{equation}
\label{lcol}
 {\cal L}_{\mbox{coll}}
   =-\frac{\hat{m}_3(\pi_0^2+\vec{\sigma}^2)}{2G\Delta_3}
    +\frac{\hat{m}_0\sigma_0}{G}
    +{\cal L}_{S}.
\end{equation}   
Since we are looking for solutions corresponding to the real vacuum state, the 
linear term $\sim\sigma_0$ in the Lagrange density (\ref{lcol}) must not occur;
it would alter the Schwinger-Dyson equations. Let us use the freedom inherent 
in the choice of the chiral symmetric piece ${\cal L}_{S}$ of 
${\cal L}_{\mbox{coll}}$ to eliminate the linear scalar field in favor of 
terms quadratic in the meson fields. There is only one chiral invariant 
combination which is bilinear in the meson fields and contains the necessary 
part with $\sigma_0$, it is $a(\sigma^2_0+\vec{\pi}^2+2\Delta_0\sigma_0 )$. The
constant $a$ is then fixed by the requirement of cancellation of the linear
term in (\ref{lcol}): $a=-\hat{m}_0(2G\Delta_0)^{-1}$. As a result, we obtain 
the solution of chiral WT identities for the bosonized NJL Lagrangian in the 
form: ${\cal L}_{\mbox{coll}}={\cal L}_{SB}+{\cal L}_{S}$, where
\begin{equation}
\label{lagsb}
 {\cal L}_{SB}=
         -\frac{\hat{m}_3(\pi_0^2+\vec{\sigma}^2)}{2G\Delta_3}
         -\frac{\hat{m}_0(\sigma_0^2+\vec{\pi}^2)}{2G\Delta_0}.
\end{equation}   
We have achieved our stated aim of integrating Eq.(\ref{sbcoll}). Let us 
turn now to the direct calculation of the fermion determinant in 
Eq.(\ref{seff}). Contrary to the chiral symmetric case, for which the 
proper-time method yields a correct result, we still have to find the way to
interpret the fermion determinant for the present problem with explicit chiral
and isospin symmetry breaking.

\section{Dirac fermion determinant with isospin symmetry breaking}

To evaluate the real part of the fermion determinant we begin with the standard
proper-time representation (\ref{logdet}). Although this definition fails when
chiral symmetry is explicitly broken, it still can be used as a basis for 
systematic calculations, if one includes the correcting counterterms in the 
functional $P$. We have also to note that in the present case the standard 
proper-time expansion does not work. There are several reasons for that. First,
it is the non-commutativity of the quark mass matrix in the heat kernel 
exponent.
This property is reflected in resummations inside the standard proper-time 
series. We denote them by T-resummations and show how to organize them. Second, 
this T-resummations do not lead automatically to chiral invariant groupings at
each order of the P-exponent. One has to search for the invariants which 
replace the Seeley-DeWitt coefficients in the asymptotic expansion of the heat 
kernel. In practice, this is the most difficult task and implies additional 
resummations. Third, the gap-equations work against chiral symmetry. One can 
restore chiral symmetry without afflicting the gap-equations, by adding to the 
Lagrangian a counterterm local in the collective fields. This is the general 
phylosophy we shall adhere to in the remainder of this paper. As a result, we 
obtain the collective Lagrangian, ${\cal L}_{\mbox{coll}}$, describing the 
low-energy limit of the four quark NJL dynamics, which is in full agreement 
with the WT identities.

\subsection{Proper-time representation for the heat kernel}

The starting point of our analysis in this section is a non-perturbative
definition of the fermion path integral, or fermion determinant:
\begin{equation}
\label{fea}
   \int {\cal D}q {\cal D}\bar{q} \exp\left\{ i\int d^4x\bar{q}Dq\right\}
   =e^{iW[\pi ,\sigma ]}\equiv \det D. 
\end{equation}
The functional $W[\pi ,\sigma ]=-i\log\det D$ is the fermionic effective action
and $D$ is given by Eq.(\ref{D}). Of course the fermion determinant defined by 
Eq.(\ref{fea}) is only a formal construction. We shall see below that $W[\pi 
,\sigma ]$ can be defined unambiguously up to a set of local counterterms on 
the basis of the modified Schwinger proper-time representation. We study here 
only the real part of the Dirac fermion determinant. 

Once a definition for $\det D$ is chosen satisfying some set of consistency 
requirements, its dependence on the collective meson fields and in particular 
its behavior under chiral transformations can be analyzed. The naive 
identification of the real part of this determinant with the proper-time 
formula Eq.(\ref{logdet}) is problematic for a couple of reasons. First, 
although the Dirac operator $D$ does not include the current quark mass, 
$\hat{m}$, the transformation law of pseudoscalar and scalar fields does. 
Thus\footnote{For the actual calculations we perform a Wick rotation into 
Euclidean space-time: $ix_0=x_4,\ i\gamma_0=\gamma_4,\ \vec{\gamma}=\gamma_i,\ 
\{\gamma_r,\gamma_s\}=-\delta_{rs}$, where $r,s=1,2,3,4$. We have chosen all 
$\gamma_r$-matrices anti-Hermitian: $\gamma_r^\dagger =-\gamma_r$. So we get 
$D\rightarrow D_{E}=i\gamma_r\partial_r-m-\sigma +i\gamma_5\pi$, where 
$\gamma_5^\dagger =\gamma_5,\ \partial_r=(\vec{\nabla}, \partial_4 )$.},
\begin{equation}
   \delta D_{E}=i[\alpha ,D_{E}]
           -i\{\gamma_5\beta ,D_{E}\}-2i\gamma_5[\hat{m}_0\beta
           +\hat{m}_3(\beta_3\tau_0 -i\gamma_5\alpha_i\epsilon_{i3k}\tau_k)],
\end{equation}
i.e., the transformation law of the Dirac operator has inhomogeneous terms
which are proportional to $\hat{m}_0$ and $\hat{m}_3$. As a result we have
\begin{eqnarray}
\label{dddd}
   \delta (D^{\dagger}_{E}D_{E})&=&
   i[\alpha +\gamma_5\beta ,D^\dagger_{E} D_{E}] \nonumber \\
   &-&4\hat{m}_0\left\{\vec{\beta}\vec{\pi}+\left[\gamma_5\epsilon_{ijk}
      (\sigma_i+m_i)\beta_j+\beta_k\pi_0\right]\tau_k\right\}  \nonumber \\
   &-&4\hat{m}_3\left\{\gamma_5\left(\alpha\pi_3-\vec{\alpha}\vec{\pi}\tau_3
      \right)+\pi_0\beta_3 \right.\nonumber \\
   &+&\left.\epsilon_{3ik}\alpha_i\sigma_k 
      +\left[\beta_3\pi_k +(\sigma_0+m_0)\epsilon_{3ik}\alpha_i\right]\tau_k
      \right\}.
\end{eqnarray}
It means that $\delta\ln |\det D|$ is not equal to zero and has systematic
contributions proportional to the current quark mass terms $\hat{m}_0$ and 
$\hat{m}_3$. This is similar to the case which has already been considered in 
\cite{Osipov:2000b}. The explicit symmetry breaking terms have to be corrected
by the corresponding contributions from the functional $P$ to lead to a
result being in agreement with the WT identities. Second, there is a new type of
contributions. These ones emerge after spontaneous symmetry breakdown and are
proportional to the isospin symmetry breaking terms $\sim (m_d-m_u)$. 
Let us discuss these problems and their solution in full detail.

We start by defining the real part of the fermion determinant by the formula:
\begin{equation}
\label{relndet}
   \mbox{Re}\left(\ln\det D\right) = \ln |\det D| + P.
\end{equation}
The first term here is the Schwinger proper-time regularization for the Dirac
fermion determinant given by Eq.(\ref{logdet}). In the case of 
nonrenormalizable models like NJL we have to introduce the cutoff $\Lambda$ to 
render the integrals over $T$ convergent. We consider a class of 
regularization schemes which can be incorporated in the expression 
(\ref{logdet}) through the kernel $\rho (T,\Lambda^2)$. These regularizations 
allow to shift in loop momenta. A typical example is the covariant 
Pauli-Villars cutoff \cite{Pauli:1949}
\begin{equation}
\label{cc}
      \rho (T, \Lambda^2)=1-(1+T\Lambda^2)e^{-T\Lambda^2},
\end{equation}
which we use in the present calculations. Our strategy is the following: 
knowing that the first term in Eq.(\ref{relndet}) breakes chiral symmetry in a 
contradictory way, we find the functional $P[\pi ,\sigma ]$ by requiring the
real part of the fermion determinant to transform in accordance with 
Eq.(\ref{sbcoll}). The WT identities are used for that. The first step in the
realization of this program is the evaluation of the heat kernel 
$\mbox{Tr}(e^{-TR})$ with the operator $R=m^2+B$, where 
\begin{equation}
\label{B}
   B=-\partial_r^2+i\gamma_r\left(\partial_r\sigma -i\gamma_5\partial_r\pi
     \right)+\sigma^2+\pi^2+\{\sigma ,m\}+i\gamma_5[\pi ,\sigma +m].
\end{equation}
We denote by
\begin{equation}
\label{m2}
   m^2=K\tau_0 +M\tau_3, \quad K=\frac{m_u^2+m_d^2}{2}, \quad 
                               M=\frac{m_u^2-m_d^2}{2},
\end{equation} 
the square of the mass matrix. The operator $R$ depends on 
functions of euclidian coordinates, $x_r$, and derivatives, $\partial_r$. 
Therefore, following the abstract formalism developed in \cite{Ball:1989}, we 
regard $e^{-TR}$ as an operator $e^{-T\hat{R}}$ acting on a fictitious Hilbert 
space, so that the heat kernel $\mbox{Tr}(e^{-TR})$ reads
\begin{equation}
\label{heatker}
   \mbox{Tr}(e^{-TR})=\int d^4x \mbox{tr}<x|e^{-T\hat{R}}|x>,
\end{equation}
where $|x>$ is an eigenvector of a commuting set of Hermitian operators 
$\hat{x}_r$ such that $\hat{x}_r|x>=x_r|x>$ and $<x|y>=\delta (x-y)$. The 
Hermitian operators, $\hat{p}_r=-i\partial_r$,  which are conjugate to 
$\hat{x}_r$, obey canonical commutation relations: $[\hat{x}_r,\hat{p}_s]=i
\delta_{rs},\ [\hat{x}_r,\hat{x}_s]=[\hat{p}_r,\hat{p}_s]=0$. Let us take an
eigenket $|p>$ of $\hat{p}_r$. Its representative in the Schr\"{o}dinger 
representation is $<x|p>=(2\pi )^{-2}\exp (ix_rp_r)$. Using this plane wave
basis one can evaluate the heat kernel directly:
\begin{equation}
   <x|e^{-T\hat{R}}|x>=\int d^4p <x|e^{-T\hat{R}}|p><p|x>.
\end{equation}
Taking into account the relations: $<x'|\partial_r|p>=\partial_r'<x'|p>$ and
$<x'|\hat{x}_r|p>=x_r'<x'|p>$ one can proceed
\begin{eqnarray}
   <x|e^{-T\hat{R}}|x>&=&\int d^4p <p|x> e^{-T\hat{R}}<x|p> \nonumber \\
   &=&\int\frac{d^4p}{(2\pi )^4}e^{-ipx}e^{-T\hat{R}}e^{ipx}.
\end{eqnarray}
We note then that $e^{-ipx}\hat{R}e^{ipx}=(\hat{R}-2ip_r\partial_r+p^2)1$,
where $p_r\partial_r=p\partial$, implies
\begin{eqnarray}
\label{op}
   <x|e^{-T\hat{R}}|x>
   &=&\int\frac{d^4p}{(2\pi )^4}e^{-Tp^2}e^{-T(\hat{R}-2ip\partial )}1 
      \nonumber \\   
   &=&\int\frac{d^4p}{(4\pi^2T)^2}e^{-p^2}
      \exp\left[-T\left(\hat{R}-\frac{2ip\partial}{\sqrt{T}}\right)\right]1.
\end{eqnarray}
Finally, combining (\ref{heatker}) and (\ref{op}), results in the desired
relation
\begin{equation}
\label{hk}
   \mbox{Tr}(e^{-TR})=\int\frac{d^4pd^4x}{(4\pi^2T)^2}e^{-(p^2+TK)}
   \mbox{tr}\left\{\exp\left[-T\left(M\tau_3+A
            \right)\right]1\right\},
\end{equation}
where
\begin{equation}
   A=B-\frac{2ip\partial}{\sqrt{T}}.
\end{equation}
Similarly as in the case of equal quark masses, one has to separate from the
heat kernal exponent the $m^2$ piece. However, in the present case $m^2$ 
contains apart from the isoscalar, $K$, which we already factorized in 
Eq.(\ref{hk}), also the non-commuting isovector term proportional to $M$. This 
non-commutativity is one of the sources of necessary $T$-resummations in the 
heat kernel expansion, which deviate from the standard case and which we 
consider next.

\subsection{P-exponent and non-commutativity of the mass matrix}

We discuss now further the exponent, $\mbox{tr}[e^{-T(M\tau_3+A)}]$, contained 
in the heat kernel. Here we can not use the standard asymptotic expansion in 
powers of the proper-time $T$, which would lead finally to the Seeley -- DeWitt
coefficients. One has first to separate the quark mass part $e^{-TM\tau_3}$, 
otherwise it will generate linear terms in $\sigma$ at any power of $T$. It is
not convenient. We need a method which leads to the gap-equations already at 
the first step of the asymptotic expansion. This separation of the
non-commuting part of the squared mass matrix in the heat kernel require the
use of the following operator identity:
\begin{equation}
\label{factor}
   e^{-T(M\tau_3+A)}=e^{-TM\tau_3}\times\mbox{P}\left\{\exp\left(-\int_0^T ds 
                     A(s)\right)\right\},
\end{equation}
where
\begin{equation}
\label{As}
   A(s)=e^{sM\tau_3}Ae^{-sM\tau_3}.
\end{equation} 
The P-exponent above is defined to mean
\begin{eqnarray}
\label{pexp}
 & &\mbox{P}\left\{\exp\left(-\int_0^T ds A(s)\right)\right\}= \nonumber \\
 &1&+\sum^\infty_{n=1}(-1)^n\int^T_0ds_1\int^{s_1}_0ds_2\ldots
   \int^{s_{n-1}}_0ds_n A(s_1)A(s_2)\ldots A(s_n). 
\end{eqnarray}
For our purpose one needs only the first three terms from the right hand side 
of this expression, i.e.,
\begin{eqnarray}
\label{ae1}
 & &\mbox{tr}[e^{-T(M\tau_3+A)}]=\mbox{tr}\left(e^{-TM\tau_3}\left\{1-TA 
    \right.\right.\nonumber \\ 
 &+&\left.\left.\frac{T^2}{4}\left[A^2+A\tau_3A\tau_3
   +c(T)(A^2-A\tau_3A\tau_3)\right]+\ldots\right\}\right),
\end{eqnarray}             
where             
\begin{eqnarray}
\label{cT}
   c(T)&=&\frac{1}{2T^2M^2}\left(e^{2TM\tau_3}-1-2TM\tau_3\right) \nonumber \\
       &=&\frac{1}{M^2}e^{TM\tau_3}\int^M_0 d\alpha [M\cosh (T\alpha )
          -\tau_3\alpha\sinh (T\alpha )].
\end{eqnarray}
The details are given in the Appendix. It is clear now that the asymptotic
heat kernel expansion (\ref{ae1}) is modified due to systematic resummations 
generated by the terms $\sim M$. After evaluating the integrals over
4-momentum $p_r$ one obtains
\begin{eqnarray}
\label{ae2}
  \mbox{Tr}(e^{-T\hat{R}})&=&\int\frac{d^4x}{(4\pi T)^2}e^{-TK}
  \mbox{tr}\left(e^{-TM\tau_3}\left\{1-TY 
  \right.\right.\nonumber \\ 
  &+&\left.\left.\frac{T^2}{4}\left[Y^2+Y\tau_3Y\tau_3
   +c(T)(Y^2-Y\tau_3Y\tau_3)\right]+\ldots\right\}\right).
\end{eqnarray}             
Here we took into account that a Lagrangian density is defined up to total 
derivatives. The $Y$ is given by
\begin{equation}
\label{Y}
   Y=i\gamma_r\left(\partial_r\sigma -i\gamma_5\partial_r\pi\right)
    +\sigma^2+\pi^2+\{\sigma ,m\}+i\gamma_5[\pi ,\sigma +m].
\end{equation}
Finally, the modulus of the fermion determinant may thus be written as
\begin{eqnarray}
\label{mod}
   &-&\ln |\det D_E|=\frac{1}{32\pi^2}\int^\infty_0\frac{dT}{T^3}\rho
   (T,\Lambda^2)e^{-TK}\mbox{tr}\left(e^{-TM\tau_3}\left\{1-TY 
   \right.\right.\nonumber \\ 
   &+&\left.\left.\frac{T^2}{4}\left[Y^2+Y\tau_3Y\tau_3
   +c(T)(Y^2-Y\tau_3Y\tau_3)\right]+{\cal O}(Y^3)\right\}\right).
\end{eqnarray} 

\subsection{Gap-equations against the chiral symmetry}

Several points about Eq.(\ref{mod}) should be clarified right away. The first 
term does not depend on collective fields and has no interest for us. Let us 
consider the second term which is proportional to $Y$ and which we denote as 
$b_1$. To simplify this expression one has to take the integral over $T$ and 
calculate the traces. The last ones include summations over isotopic, color and
euclidean indices. The integrals over $T$ can be reduced to combinations of 
some set of elementary integrals $J_n(m^2)$
\begin{equation}
\label{Jn}
   J_n(m^2)=\int^\infty_0\frac{dT}{T^{2-n}}e^{-Tm^2}\rho (T,\Lambda^2),
            \quad n=0,1,2\ldots  .
\end{equation}
These manipulations lead us to the result:
\begin{eqnarray}
\label{b1}
   b_1&=&-\frac{N_c}{8\pi^2}\left\{[J_0(m^2_u)+J_0(m^2_d)][\sigma_a^2 
         +\pi^2_a +2(m_0\sigma_0+m_3\sigma_3 )] \right.\nonumber \\
      &+&\left. 2[J_0(m^2_u)-J_0(m^2_d)](\sigma_0\sigma_3 +\pi_0\pi_3 
         +m_0\sigma_3+m_3\sigma_0 )\right\}.
\end{eqnarray}
Other terms in the expansion of the integrand in (\ref{mod}) are $\sim Y^n$, 
where $n\ge 2$, and therefore do not contribute to the term linear in $\sigma$. 
We consider small fluctuations of the system about the asymmetric vacuum state
and have already shifted the scalar fields correspondingly. It means that linear
terms in the scalar fields should not be present in the Lagrangian. This 
self-consistency requirement can be expressed in terms of the 
gap-equations:
\begin{equation}
\label{gapu}
   \frac{m_u-\hat{m}_u}{m_u}=\frac{N_cG}{2\pi^2}J_0(m_u^2),
\end{equation}
\begin{equation}
\label{gapd}
   \frac{m_d-\hat{m}_d}{m_d}=\frac{N_cG}{2\pi^2}J_0(m_d^2).
\end{equation}
Once this has been done, one is confronted with a problem related to the fact
that the last term in Eq.(\ref{b1}), linear in the scalar fields, i.e.,
\begin{equation}
\label{diff}
   -\frac{N_c}{4\pi^2}[J_0(m^2_u)-J_0(m^2_d)](m_0\sigma_3+m_3\sigma_0)
\end{equation}
vanishes. Let us make clear the essence of the problem. The difference of two 
$J_0$ integrals with different arguments can be written as a sum of two $J_1$ 
multiplied by M and a rest which includes the contributions from $J_n$ 
integrals with $n>1$
\begin{equation}
\label{diff1}
   J_0(m^2_u)-J_0(m^2_d)=-M[J_1(m^2_u)+J_1(m^2_d)]+{\cal O}(M^3).
\end{equation}
As we soon will see, the first term contributes to the next order of our
modified proper-time expansion. This fact is very important, because without
this contribution chiral symmetry would be destroyed at the next order. The 
situation is even more complicated, since the ${\cal O}(M^3)$ terms will of
course contribute to all remaining orders of the heat kernel expansion. 
For the same reason relation (\ref{diff1}) implies also that the term quadratic 
in the fields in Eq.(\ref{b1}), multiplying the difference of $J_0$ integrals,
does not contribute to the leading order. Thus in evaluating the fermion 
determinant on the basis of representation (\ref{mod}) one has to perform 
carefully systematic resummations in the Schwinger's proper-time expansion. 
This is an embarrassment but not a catastrophy. Indeed, from formula 
(\ref{dddd}) it follows that $\delta\ln |\det D_E|\sim \hat{m}$. Therefore the 
resummations are possible in general, since the total expression (\ref{logdet})
is already a chiral quasi-invariant, i.e., invariant up to terms $\sim\hat{m}$.
However, the ansatz introduced by the gap-equations destroys this picture, 
taking away one of the necessary elements. To restore the transformation 
property of the determinant instead of the removed term (\ref{diff}) one has to
add to the functional $P$ a counterterm which is a quadratic polynomial in the 
meson fields, and which has the same transformation property as the removed 
expression (\ref{diff}). The correction predicted by the WT identities thus 
naturally takes the form 
\begin{equation}
\label{newdiff}
   \frac{N_c}{8\pi^2}[J_0(m^2_u)-J_0(m^2_d)]\frac{1}{\Delta_0\Delta_3}
   [\Delta_0^2(\vec{\sigma}^2+\pi_0^2)+\Delta_3^2(\sigma_0^2+\vec{\pi}^2)].
\end{equation}

\subsection{Extracting quasi-invariants}

Let us turn to Eq.(\ref{b1}) and consider the first term. This term together
with the second one in Eq.(\ref{seff}) gives the part of the meson Lagrangian
which is responsible for the explicit symmetry breaking. Using the 
gap-equations one obtains 
\begin{equation}
   {\cal L}_{SB}'=-\frac{1}{4G}\left(
   \frac{\hat{m}_u}{m_u}+\frac{\hat{m}_d}{m_d} \right)
   (\sigma^2_a +\pi^2_a).
\end{equation}
This result is in obvious contradiction with Eq.(\ref{lagsb}). But this should 
not surprise us because we already know that the proper-time regularization 
destroys the explicit symmetry breaking pattern of the theory. Fortunately, this
part of the bosonic Lagrangian can be corrected by the appropriate counterterm
from the polynomial $P$ which is unambiguously fixed by the chiral WT 
identities. The following circumstances do extremely simplify the problem: 

(1) The symmetry breaking part of the NJL Lagrangian must be proportional to
the $J_0$ integrals and not to any other $J_n$. This follows from the form of
the gap-equations and from the symmetry breaking pattern of the model. Thus, 
all steps in the asymptotic expansion of the heat kernel, excluding the 
first one, must be chiral symmetric.

(2) It is possible to reduce the problem of finding the explicit form of these
chiral invariants to the problem of deriving quasi-invariants, i.e.,  
groups of terms which would be chiral invariant if one would add to them the
deficient terms depending explicitly on the current quark masses. 

The first part of this program requires resummations. The second one can 
be solved by considering corresponding counterterms in the polynomial $P$. 

To see these ideas at work let us restrict ourselves to the easiest case, in 
which the first quasi-invariants will be obtained. We shall carry out the 
necessary arguments in detail only up to the second step in the asymptotic 
expansion; the generalization to the next steps will be more tedious but 
straightforward. Let us consider the third term in Eq.(\ref{mod}), quadratic in
$Y$, which we denote by $b_2$. First, one 
has to perform the $T$ integration. Using formula (\ref{cT}) for $c(T)$, and 
noting that the part of the integrand $\sim\sinh (T\alpha )$ in that expression 
does not contribute to Eq.(\ref{mod}), since the trace over isospin matrices  
vanishes, one obtains  
\begin{eqnarray}
\label{b2}
   b_2&=&\frac{1}{(16\pi )^2}\mbox{tr}\left\{
         \left[J_1(m^2_u)+J_1(m^2_d)\right]\left(Y^2+Y\tau_3Y\tau_3\right)
         \right. \nonumber \\
      &+&2\left[J_1(m^2_u)-J_1(m^2_d)\right]\tau_3Y^2
         \nonumber \\
      &+&\frac{1}{M}\left.
         \left[J_0(m^2_d)-J_0(m^2_u)\right]\left(Y^2-Y\tau_3Y\tau_3\right)
         \right\}.
\end{eqnarray}
We use now Eq.(\ref{diff1}) to recast this expression in the form
\begin{eqnarray}
\label{b2s}
   b_2&=&\frac{2}{(16\pi )^2}\mbox{tr}\left\{
         \left[J_1(m^2_u)+J_1(m^2_d)\right]Y^2
         +\left[J_1(m^2_u)-J_1(m^2_d)\right]\tau_3Y^2
         \right. \nonumber \\    
      &+&\left. Q\left(Y^2-Y\tau_3Y\tau_3\right)
         \right\},
\end{eqnarray}
where   
\begin{eqnarray} 
\label{Q}
   Q&=&\frac{1}{2M}\left\{K
       \left[J_1(m^2_u)-J_1(m^2_d)\right]
       +\frac{2M\Lambda^4}{(\Lambda^2+m_u^2)(\Lambda^2+m_d^2)}
       \right\} \nonumber \\
    &=&-\frac{M^2\Lambda^4}{K(\Lambda^2+K)^3}+\ldots\ .
\end{eqnarray}
The second term in Eq.(\ref{b2s}) contributes as $\sim M$ and the third one as
$\sim M^2$. It means that these terms contribute to the next and next to the
next orders of the asymptotic expansion. At the level of the considered 
approximation one has to take into account only the first term in 
Eq.(\ref{b2s}). The trace $\mbox{tr}(Y^2)$ is equal to
\begin{eqnarray}
\label{trY2}
   \mbox{tr}(Y^2)&=&4N_c\left\{
                 (\partial_r\sigma_a)^2+(\partial_r\pi_a)^2+
                 \left[\sigma^2_a +\pi^2_a 
                 +2(m_0\sigma_0+m_3\sigma_3)\right]^2
                 \right. \nonumber \\
    &+&4\left[\vec{\pi}^2(\sigma_i+m_i)^2-(\vec{\pi}\vec{\sigma}+m_3\pi_3)^2
     +\left(\pi_i\pi_0+\sigma_i(\sigma_0+m_0)\right)^2
     \right.\nonumber \\
    &+&\left.\left.  m_3^2\sigma_0^2
     +2m_3\sigma_0\left(\pi_0\pi_3+\sigma_3(\sigma_0+m_0)\right)
     \right]\right\}.
\end{eqnarray}
It is not difficult to recognize a quasi-invariant
$[\sigma^2_a +\pi^2_a +2(m_0\sigma_0+m_3\sigma_3)]^2$ in the first line. 
Indeed, the
chiral transformations (\ref{stbf}) leave invariant the expression 
$[\sigma^2_a +\pi^2_a +2(\Delta_0\sigma_0+\Delta_3\sigma_3)]^2$ 
which is obtained from the previous one by adding the terms proportional to the
current quark masses $\hat{m}_0$ and $\hat{m}_3$. The second and the third 
lines in Eq.(\ref{trY2}) do not have a quasi-invariant form. However at this
stage there are contributions from $b_1$ (see the term $\sim (\sigma_0\sigma_3
+\pi_0\pi_3)$ in Eq.(\ref{b1})) and from the polynomial $P$ (see the 
counterterm (\ref{newdiff})), where in the latter one has to take into 
account only the parts which contribute as $[J_1(m^2_u)+J_1(m^2_d)]$ 
(see Eq.(\ref{diff1}). As a result we obtain a quasi-invariant structure:
\begin{equation}
\label{qinv2}
   (\pi_0^2+\vec{\sigma}^2+2m_i\sigma_i)(\sigma_0^2+\vec{\pi}^2+2m_0\sigma_0)
   -\left[\pi_i(\sigma_i+m_i)-\pi_0(\sigma_0+m_0)\right]^2.
\end{equation}
We can now adjust this expression to a new one which is invariant under 
chiral transformations by including a corresponding counterterm to the 
polynomial $P$. The possible counterterm which compensates the non-zero 
variation of (\ref{qinv2}) is easily constructed through the replacement: 
$m_\alpha\rightarrow\Delta_\alpha$ directly in Eq.(\ref{qinv2}). In agreement 
with the symmetry requirement we obtain now
\begin{equation}
\label{inv2}
   (\pi_0^2+\vec{\sigma}^2+2\Delta_i\sigma_i)(\sigma_0^2+\vec{\pi}^2
   +2\Delta_0\sigma_0)
   -\left[\pi_i(\sigma_i+\Delta_i)-\pi_0(\sigma_0+\Delta_0)\right]^2
\end{equation}
instead of (\ref{qinv2}). Finally, if we collect all terms which contribute to 
the considered approximation, we obtain a Lagrange density of the bosonized NJL
model in the form
\begin{eqnarray}
\label{lagran} 
   {\cal L}_{\mbox{coll}}
   &=&-\frac{\hat{m}_3(\pi_0^2+\vec{\sigma}^2)}{2G\Delta_3}
      -\frac{\hat{m}_0(\sigma_0^2+\vec{\pi}^2)}{2G\Delta_0}
      -\frac{N_c}{16\pi^2}\left[J_1(m^2_u)+J_1(m^2_d)\right]         
      \nonumber \\
   &\times&\left\{
      (\partial_r\sigma_a)^2+(\partial_r\pi_a)^2+
      \left[\sigma^2_a +\pi^2_a 
      +2(\Delta_0\sigma_0+\Delta_3\sigma_3)\right]^2
      \right. \nonumber \\
   &+&4\left[(\pi_0^2+\vec{\sigma}^2+2\Delta_3\sigma_3)(\sigma_0^2+\vec{\pi}^2
      +2\Delta_0\sigma_0)
      \right. \nonumber \\
   &-&\left.\left.\left(\pi_i(\sigma_i+\Delta_i)-\pi_0(\sigma_0+\Delta_0)
      \right)^2\right]\right\}.   
\end{eqnarray}
In closing, we remark that the method developed so far allows for an accurate
derivation of the bosonized NJL Lagrangian which describes the low-energy
dynamics of Goldstone bosons coupled with scalar fields. This Lagrangian
consistently summarizes the effect of the isospin symmetry breaking, $m_u\ne 
m_d$, and allows for a calculation of its physical consequences. The expression 
(\ref{lagran}) fulfills the chiral symmetry requirements of the fundamental NJL
quark Lagrangian permitting us to consider it as an effective low-energy
approximation to the chiral quark dynamics described by the Lagrange density
(\ref{enjl}).

\section{Coupling constants and masses}

For completeness, we would like to discuss here the coupling constants and 
masses of the physical collective modes in the Lagrangian (\ref{lagran}). We 
need to bring this expression to the standard form, i.e., to diagonalize its
bilinear part and to introduce the physical states by the corresponding field
renormalizations. By rewriting the Lagrangian in terms of physical states we 
complete our work in its analytical part. We are not going to give here the 
numerical discussion of the resulting mass formulas for composite mesons or to 
fix the parameters of the model. This numerical part is not relevant for our
consideration here and will be done elsewhere together with the discussion of
some physical problems.   

Let us consider the part of the Lagrange density (\ref{lagran}) which is 
quadratic in the fields 
\begin{eqnarray}
\label{freelag}
   {\cal L}_{\mbox{free}}\!
   &=&\!-\frac{\hat{m}_3(\pi_0^2+\vec{\sigma}^2)}{2G\Delta_3}
      -\frac{\hat{m}_0(\sigma_0^2+\vec{\pi}^2)}{2G\Delta_0}
      -\frac{I}{4}\left[(\partial_r\sigma_a)^2
                       +(\partial_r\pi_a)^2\right]
      \nonumber \\
   &-&\! I\left[ \Delta^2_0(\sigma^2_0-\pi^2_0)
      +\Delta^2_3(\sigma^2_3-\pi^2_3)
      +2\Delta_0\Delta_3(\pi_0\pi_3+3\sigma_0\sigma_3)
      \right],
\end{eqnarray}
where
\begin{equation}
   I=\frac{N_c}{4\pi^2}\left[J_1(m^2_u)+J_1(m^2_d)\right].         
\end{equation}
There are two kinds of mixing: $\sigma_0$-$\sigma_3$ and $\pi_0$-$\pi_3$. Both 
of them arise as a result of isospin symmetry breaking, and both are 
diagonalized by the orthogonal rotation:
\begin{equation}
   {\phi_0\choose\phi_3}={\cos\theta\ \ \sin\theta \choose -\sin\theta\ \ 
   \cos\theta}{\tilde{\phi}_0\choose\tilde{\phi}_3},
\end{equation}
where we have chosen the general notation $\phi$ which means either $\sigma$ 
or $\pi$, depending on the case under consideration. The rotation angle 
specified by the $\theta$'s is equal to $\theta_s$ for scalars and to
$\theta_p$ for pseudoscalars. The quadratic form which has to be diagonalized 
is $-(A\phi_0^2+C\phi_0\phi_3+B\phi_3^2)$. Then the angle of the rotation is 
fixed by the condition: $\tan (2\theta )=C/(B-A)$. From Eq.(\ref{freelag}) one 
obtains
\begin{equation}
     \tan 2\theta_s =-\frac{6\Delta_0\Delta_3}{\omega}, \quad
     \tan 2\theta_p =\frac{2\Delta_0\Delta_3}{\omega}, 
\end{equation}
where
\begin{equation}
     \omega =\Delta_0^2-\Delta_3^2+\frac{1}{2GI}
             \left(\frac{\hat{m}_0}{\Delta_0}-
             \frac{\hat{m}_3}{\Delta_3}\right).
\end{equation} 
For convenience let us rename $\phi_{1,2}=\tilde{\phi}_{1,2}$. Having introduced
the relevant fields, we are now in position to write down the free Lagrangian 
(\ref{freelag}) as 
\begin{eqnarray}
\label{freelag2}
   {\cal L}_{\mbox{free}}
   &=&-\frac{\hat{m}_3(\tilde{\sigma}_1^2+\tilde{\sigma}_2^2)}{2G\Delta_3}
      -\frac{\hat{m}_0(\tilde{\pi}_1^2+\tilde{\pi}_2^2)}{2G\Delta_0}
      -\frac{I}{4}\left[
       (\partial_r\tilde{\sigma}_a)^2+(\partial_r\tilde{\pi}_a)^2
       \right]     \nonumber \\
   &-&\left(\tilde{A}_s\tilde{\sigma}_0^2+\tilde{B}_s\tilde{\sigma}_3^2
      +\tilde{A}_p\tilde{\pi}_0^2+\tilde{B}_p\tilde{\pi}_3^2\right).
\end{eqnarray}
The coefficients with tilde are defined as
\begin{equation}
   \tilde{A}=A\cos^2\theta +B\sin^2\theta -C\sin\theta\cos\theta ,
\end{equation}
\begin{equation}
   \tilde{B}=A\sin^2\theta +B\cos^2\theta +C\sin\theta\cos\theta .
\end{equation}
with the following values of coefficients $A$, $B$, and $C$ for the scalar and
pseudoscalar cases respectively (see Eq.(\ref{freelag})) 
$$
  A_s=\frac{\hat{m}_0}{2G\Delta_0}+I\Delta_0^2, \quad
  B_s=\frac{\hat{m}_3}{2G\Delta_3}+I\Delta_3^2, \quad
  C_s=6I\Delta_0\Delta_3, 
$$
\begin{equation}
  A_p=\frac{\hat{m}_3}{2G\Delta_3}-I\Delta_0^2, \quad
  B_p=\frac{\hat{m}_0}{2G\Delta_0}-I\Delta_3^2, \quad
  C_p=2I\Delta_0\Delta_3.
\end{equation}
The physical states are defined by bringing the kinetic terms in the ${\cal 
L_{\mbox{free}}}$ to the standard form, i.e.,
\begin{equation}
   \tilde{\sigma}_\alpha =g\sigma_\alpha^{(ph)}, \quad
   \tilde{\pi}_\alpha =g\pi_\alpha^{(ph)}, \quad Ig^2=2.
\end{equation}   
Finally these replacements determine the masses to be
\begin{equation}
   m^2_{\sigma_{1,2}}=\frac{\hat{m}_3g^2}{G\Delta_3}, \quad 
   m^2_{\sigma_3}=2g^2\tilde{B}_s, \quad
   m^2_{\sigma_0}=2g^2\tilde{A}_s. 
\end{equation}
\begin{equation}
   m^2_{\pi_{1,2}}=\frac{\hat{m}_0g^2}{G\Delta_0}, \quad 
   m^2_{\pi_3}=2g^2\tilde{B}_p, \quad
   m^2_{\pi_0}=2g^2\tilde{A}_p, 
\end{equation}
They are different, for example, from the results obtained in 
\cite{Volkov:1984} and \cite{Ebert:1986,Bijnens:1996}. They differ also from 
the ones obtained on the basis of direct calculations with Feynman diagrams 
\cite{Klevansky:1992,Hatsuda:1994} (and references therein). This can be easily
understood, recalling the general symmetry properties of the model discussed in
section 2: the chiral transformations of meson fields depend explicitly on the 
current quark masses, see Eq.(\ref{stbf}), but the Dirac operator, (\ref{D}) 
does not. Therefore the Feynman amplitudes calculated directly and solely from 
$\ln\det D$ will never depend on the current quark masses and consequently not 
fulfill the symmetry requirements. To have the correct behavior one needs 
chiral invariant combinations of fields, i.e. the missing current quark mass 
terms have to be included in the correcting polynomial $P$, via the WT 
identities, as extensively discussed in this paper.  

\section{Concluding remarks}

The mesonic degrees of freedom $\sigma_\alpha$, and $\pi_\alpha$ are the 
elementary collective excitations of the non-trivial ground state of the NJL 
model. The chiral dynamics of this collective modes should be consistent with
the standard picture of a spontaneously broken chiral symmetry, which is
naturally linked to the explicit symmetry breaking pattern of the underlying 
four-quark Lagrangian. In this paper we studied the NJL model with the broken 
chiral and diagonal $U(2)$ flavour symmetry, choosing the quark mass matrix to 
be in the non-degenerate form: $\hat{m}_u\ne\hat{m}_d$. We suggested and 
discussed in considerable detail a new method to derive the effective bosonic 
Lagrangian starting from the pure quark action with the chiral four fermion 
interactions. The collective bosonic modes have been introduced through the 
generating functional. It is known that quite often the identification of the 
path integral over quark fields with a determinant is problematic. The case 
under consideration was not an exclusion. The method we developed deals with 
the accurate definition and evaluation of the low energy part of the fermion 
determinant. We studied the real part of the quark effective action and 
obtained the low energy expansion for it. To get the expansion we have chosen 
the Schwinger's proper-time representation of the fermion determinant as a 
starting point. We have shown that in this case one can not use the standard 
proper-time expansion. The mass matrix of quark fields, which does not commute 
with the mesonic part of the Dirac operator, has first to be factorized from 
the heat kernel. We used the time-ordered exponent to separate the part of the 
heat kernel which depends on the meson fields. In this case the expansion of 
the time-ordered exponent leads directly to the gap-equations. 

On the basis of our recent results \cite{Osipov:2000a,Osipov:2000b} we knew
that a proper-time representation failed in the case of manifestly broken 
chiral symmetry. Therefore we used the WT identities to correct the 
$\hat{m}$-dependent part of the fermion determinant. In this part of the work 
we extended the method developed in our previous papers to the isospin
asymmetric case. However an additional new problem arose: we found that the 
naive expansion of the time-ordered exponent has to be modified by additional 
resummations which touch the terms of the effective action which are 
proportional to the difference of constituent quark masses $m_d-m_u$. These 
resummations are necessary to form groups of terms which are invariant under 
the chiral transformations. 

The gap-equation ansatz made the picture even more intricate. We have found 
that because of resummations the contributions of linear terms in scalar fields
are important for chiral symmetry. As a consequence one has to add the 
correcting counterterm to the effective action to put the ansatz in agreement 
with the WT requirements. 

The method presented in this work is a necessary mathematical tool which can be
applyed to the more interesting, from the physical point of view, case of 
explicitly broken chiral $SU(3)\times SU(3)$ symmetry. These versions of the 
NJL model have been already studied but without a careful treatment of the 
questions indicated in this paper. 

\section*{Acknowledgements}

This work is supported by grants provided by Funda\ca o para a Ci\^encia e a
Tecnologia, PRAXIS/C/FIS/12247/1998, PESO/P/PRO/15127/1999, 
POCTI/1999/FIS/35304 and NATO "Outreach" Cooperation Program.
 
\section*{Appendix}

Let us obtain here the first four terms in the asymptotic long-wave expansion 
of the operator
\begin{equation}
   \mbox{tr}\left[e^{-T(M\tau_3+A)}\right]. 
\end{equation}
This operator is a part of the heat kernel in the expession (\ref{hk}). The
result, as we already know, has been used in Eq.(\ref{ae1}). We begin here by 
making use of the well known formalism of the ``time"-ordered exponent, 
or $P$ exponent:
\begin{equation}
\label{Mt3A}
   \mbox{tr}\left[e^{-T(M\tau_3+A)}\right]
   =\mbox{tr}\left\{e^{-T(M\tau_3)}P\left[
   \exp\left(-\int^T_0 ds A(s)\right)\right]
   \right\}=\sum_{i=0}^{\infty}A_i,  
\end{equation}
where the $P$-exponent is defined by Eq.(\ref{pexp}), and $A(s)$, according to
the definition (\ref{As}), is equal to 
\begin{equation}
   A(s)=e^{sM\tau_3}Ae^{-sM\tau_3}=\frac{1}{2}\left[A+\tau_3A\tau_3
        +e^{2sM\tau_3}\left(A-\tau_3A\tau_3\right)\right]. 
\end{equation}         
One can easily verify that the second term of this expression does not 
contribute to the first nontrivial term of the $P$ exponent, for its isotopic 
trace vanishes. Thus the first two terms in Eq.(\ref{ae1}) are obvious. 
Let us consider now the integral
\begin{eqnarray}
\label{intAA}
  & &\int^T_0 ds \int^s_0 ds_1 A(s)A(s_1) \nonumber \\
  &=&\frac{1}{4M}\int^T_0 ds \left[ 
     sM(A^2+A\tau_3A\tau_3+\tau_3A\tau_3A+\tau_3A^2\tau_3) 
     \right. \nonumber \\ 
  &+&sMe^{2sM\tau_3}(A^2+A\tau_3A\tau_3-\tau_3A\tau_3A-\tau_3A^2\tau_3)
     \nonumber \\
  &+&\left. 2\sinh (sM)e^{sM\tau_3}(A^2-A\tau_3A\tau_3)\right].
\end{eqnarray}
The explicit form for the third term in the expansion of Eq.(\ref{Mt3A}) is
given by
\begin{eqnarray}
\label{A2}
 A_2&=&\mbox{tr}\left\{\frac{1}{2M}e^{-TM\tau_3} \int^T_0 ds 
       \left[sM(A^2+A\tau_3A\tau_3) \right.\right. \nonumber \\
    &+&\left.\left. \sinh (sM) e^{sM\tau_3}(A^2-A\tau_3A\tau_3)\right]\right\}.
\end{eqnarray}
Note that the second term in Eq.(\ref{intAA}) does not contribute to $A_2$,
by the property of the trace, which vanishes. If we integrate over $s$, we find
from (\ref{A2}) the third term in the expansion (\ref{ae1}). The calculation
of the coefficients $A_i$ becomes a somewhat more and more difficult exercise at
every new step in (\ref{Mt3A}). To finish this Appendix we give the result of
such calculations for the coefficient $A_3$:
\begin{eqnarray}
  A_3&=&-\frac{T^3}{8}\mbox{tr}\left\{e^{-TM\tau_3}\left[
         \frac{A}{3}\{A,\tau_3\}^2+c_2(T)A\{A,\tau_3\}[\tau_3,A]
         \right.\right. \nonumber \\
     &+&\left.\left. c_3(T)(A^3-A\tau_3A\tau_3A)
         \right]\right\},
\end{eqnarray}
where the matricies $c_2(T)$ and $c_3(T)$ may be written as
\begin{equation}
   c_2(T)=\frac{\tau_3}{2T^3M^3}\left[1+TM\tau_3+
          (TM\tau_3-1)e^{2TM\tau_3}\right],
\end{equation}
\begin{equation}
   c_3(T)=\frac{\tau_3}{2T^3M^3}\left[e^{2TM\tau_3}-1-2TM\tau_3-2T^2M^2
          \right].
\end{equation}
                                                         
\baselineskip 12pt plus 2pt minus 2pt

\end{document}